\documentclass[conference]{IEEEtran}
\IEEEoverridecommandlockouts
\usepackage{graphicx} 
\usepackage{tabularx}
\usepackage{booktabs} 
\usepackage{array}
\usepackage{url}
\usepackage{tikz}
\usetikzlibrary{arrows.meta, positioning, fit}

\title{Dataflows and Computational Patterns for Hybrid Quantum-Classical Scientific Computing}
\author{\IEEEauthorblockN{Ryan Landfield}
\IEEEauthorblockA{\textit{Oak Ridge National Laboratory} \\
Oak Ridge, TN, USA \\
landfieldre@ornl.gov}
\and
\IEEEauthorblockN{Jordan J. Winetrout}
\IEEEauthorblockA{\textit{Oak Ridge National Laboratory} \\
Oak Ridge, TN, USA \\
winetroutjj@ornl.gov}
\and
\IEEEauthorblockN{Michael A. Sandoval}
\IEEEauthorblockA{\textit{Oak Ridge National Laboratory} \\
Oak Ridge, TN, USA \\
sandovalma@ornl.gov}
}

\begin{document}

\maketitle

\begin{abstract}
Hybrid quantum-classical computing has emerged as the dominant paradigm for near-term quantum applications, yet hybrid workflows are typically described by individual algorithms rather than their underlying execution behavior. We introduce the \emph{Quantum Execution Locality Framework (QELF)}, a qualitative framework for characterizing hybrid quantum-classical workflows according to recurring dataflow structures and quantum execution locality, the extent to which computation remains resident on the Quantum Processing Unit (QPU) before host intervention or classical synchronization. From a representative cross-section of applications, QELF identifies five recurring computational patterns with distinct locality characteristics and discusses their implications for communication overhead, workflow organization, and future hybrid computing architectures. By providing a common vocabulary for reasoning about hybrid workloads, QELF establishes a foundation for future quantitative validation and the co-design of algorithms, runtime systems, and hybrid computing architectures.
\end{abstract}


\section{Introduction}

The evolution of scientific computing has repeatedly been driven by new accelerator architectures and the reformulation of applications to exploit them. Graphics Processing Units (GPUs), for example, achieved widespread adoption by maximizing computational locality through large, contiguous accelerator regions while minimizing host-device communication~\cite{nickolls2008cuda}.

Hybrid quantum-classical computing presents a similar challenge. Today's applications typically embed short quantum kernels within larger classical workflows, resulting in frequent synchronization, data movement, and orchestration overhead between High Performance Computing (HPC) systems and Quantum Processing Units (QPUs)~\cite{Beck2024IntegratingQuantumHPC}. While effective for current noisy quantum devices, these fragmented workflows may limit the long-term potential of quantum acceleration.

To address this challenge, we introduce the \emph{Quantum Execution Locality Framework (QELF)}, which characterizes hybrid workflows according to recurring dataflow structures and quantum execution locality. Rather than grouping algorithms by scientific domain, QELF groups applications exhibiting similar host--QPU communication patterns, synchronization cadence, and opportunities for sustained quantum execution. Quantum execution locality is defined as the extent to which computation remains resident on the QPU before host intervention or classical synchronization.

Using QELF, we classify representative hybrid workflows into recurring computational patterns and examine how locality influences communication overhead, hardware utilization, and opportunities for sustained quantum acceleration. We argue that an equally important question is not only which problems admit quantum algorithms, but how scientific workflows can be reformulated to increase quantum-resident computation.

\section{Hybrid Dataflows \& Locality}
We represent hybrid quantum-classical applications as directed dataflows in which nodes correspond to computation executed on either classical HPC resources or QPUs, while edges represent the movement of data and control between them. This abstraction is intentionally independent of programming models or hardware implementations, allowing diverse applications to be compared using a common structural framework.

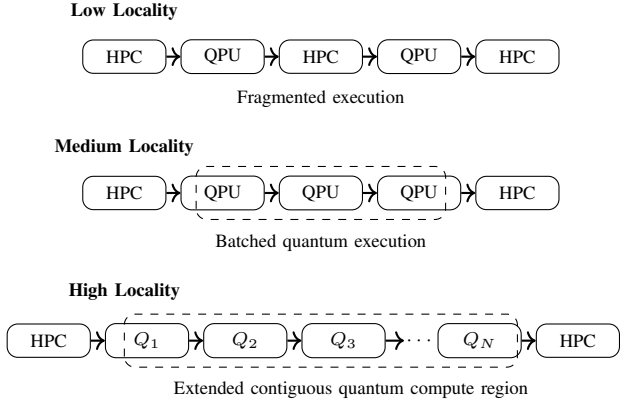
\begin{figure}[t]
\centering
\begin{tikzpicture}[
    node distance=0.55cm,
    every node/.style={font=\scriptsize},
    cnode/.style={draw, rounded corners, align=center, minimum width=1.1cm, minimum height=0.44cm},
    qnode/.style={draw, rounded corners, align=center, minimum width=1.1cm, minimum height=0.44cm},
    arrow/.style={->, thick}
]

\node at (0,1.2) {\textbf{Low Locality}};
\node[cnode] (c1) at (0,0.6) {HPC};
\node[qnode] (q1) at (1.3,0.6) {QPU};
\node[cnode] (c2) at (2.6,0.6) {HPC};
\node[qnode] (q2) at (3.9,0.6) {QPU};
\node[cnode] (c3) at (5.2,0.6) {HPC};
\draw[arrow] (c1) -- (q1);
\draw[arrow] (q1) -- (c2);
\draw[arrow] (c2) -- (q2);
\draw[arrow] (q2) -- (c3);
\node at (2.6, 0.05) {Fragmented execution};
\node at (0,-0.6) {\textbf{Medium Locality}};
\node[cnode] (m1) at (0,-1.2) {HPC};
\node[qnode] (mq1) at (1.3,-1.2) {QPU};
\node[qnode] (mq2) at (2.6,-1.2) {QPU};
\node[qnode] (mq3) at (3.9,-1.2) {QPU};
\node[cnode] (m2) at (5.2,-1.2) {HPC};
\draw[arrow] (m1) -- (mq1);
\draw[arrow] (mq1) -- (mq2);
\draw[arrow] (mq2) -- (mq3);
\draw[arrow] (mq3) -- (m2);
\draw[dashed, rounded corners] (0.95,-1.55) rectangle (4.25,-0.85);
\node at (2.6,-1.85) {Batched quantum execution};
\node at (0,-2.5) {\textbf{High Locality}};

\node[cnode] (h1) at (-1.0,-3.15) {HPC};

\node[qnode] (hq1) at (0.3,-3.15) {$Q_1$};
\node[qnode] (hq2) at (1.6,-3.15) {$Q_2$};
\node[qnode] (hq3) at (2.9,-3.15) {$Q_3$};

\node at (3.9,-3.15) {$\cdots$};

\node[qnode] (hqn) at (4.7,-3.15) {$Q_N$};

\node[cnode] (h2) at (6.0,-3.15) {HPC};

\draw[arrow] (h1) -- (hq1);
\draw[arrow] (hq1) -- (hq2);
\draw[arrow] (hq2) -- (hq3);
\draw[arrow] (hq3) -- (3.7,-3.15);
\draw[arrow] (hqn) -- (h2);

\draw[dashed, rounded corners]
    (0.0,-3.5) rectangle (5.2,-2.8);

\node at (3.0,-3.8)
    {Extended contiguous quantum compute region};
\end{tikzpicture}
\caption{Locality regimes in hybrid quantum-classical dataflows. Low-locality workflows repeatedly alternate between classical and quantum execution, medium-locality workflows batch quantum operations before classical synchronization, and high-locality workflows contain contiguous quantum compute regions in which computation remains resident on the QPU.}
\label{fig:locality-spectrum}
\end{figure}

Within these dataflows, the most important characteristic is quantum execution locality, the extent to which computation remains resident on the QPU before measurement or synchronization with classical resources. Figure~\ref{fig:locality-spectrum} illustrates a contiguous quantum compute region, in which quantum data undergoes multiple transformations without intermediate offloading. Such regions are analogous to GPU kernels in heterogeneous HPC, where sustained on-device execution minimizes communication overhead and maximizes accelerator utilization. 

Using this perspective, hybrid workflows naturally fall into three broad classes: 

\begin{enumerate}
    \item Fragmented (low locality) -- repeated classical-quantum interaction separates short quantum kernels.
    \item Batched (medium locality) -- groups of quantum operations execute before classical synchronization.
    \item Contiguous (high locality) -- extended quantum-resident computation minimizes interaction with the classical system.
\end{enumerate}

This locality-based view, shown in Figure~\ref{fig:locality-spectrum}, provides a common basis for comparing hybrid workflows independent of application domain. As opposed to classifying applications solely by scientific discipline or quantum algorithm, we characterize them by their underlying dataflow structure, enabling recurring computational patterns to be identified across diverse areas of hybrid scientific computing.

\section{Hybrid Computational Pattern Taxonomy}

The computational patterns proposed in this work emerged from a structural analysis of representative hybrid quantum-classical applications spanning quantum chemistry, materials science, optimization, machine learning, linear algebra, and scientific simulation \cite{yu2025skqd} \cite{cerezo2021variational} \cite{peruzzo2014variational} \cite{harrow2009hhl} \cite{lloyd1996universal} \cite{montanaro2015montecarlo} \cite{brassard2002amplitude}. Rather than grouping algorithms by their mathematical formulation, the proposed patterns are derived from the structure of data movement and synchronization between classical and quantum resources. Applications are grouped when they exhibit similar host-QPU communication structure, synchronization cadence, and opportunities for sustained quantum execution, irrespective of scientific domain. This perspective exposes common dataflow structures across diverse applications and provides a natural basis for comparing quantum execution locality.

The five primary computational patterns identified in this work are summarized in Table~\ref{tab:patterns}.

\begin{itemize}
    \item \textbf{QOOL -- Quantum Optimization Outer Loop:} A classical optimizer repeatedly invokes a quantum kernel to evaluate an objective function (e.g., VQE, QAOA).

    \item \textbf{QASP -- Quantum-Assisted Subspace Projection:} The QPU generates basis states or projected quantities that are assembled and solved classically (e.g., Sample-based Krylov Diagonalization).

    \item \textbf{QSMC -- Quantum Sampling / Monte Carlo:} The QPU generates samples from complex probability distributions that are subsequently aggregated and analyzed using classical resources.

    \item \textbf{QTES -- Quantum Time Evolution / Simulation:} The QPU performs extended quantum-resident time evolution or simulation before measurement and classical analysis.

    \item \textbf{QLAO -- Quantum Linear Algebra Oracle:} The QPU serves as a solver for linear algebra primitives such as eigenvalue estimation or linear system solution, with classical resources performing problem setup and post-processing.
\end{itemize}

Although these patterns span a broad range of scientific applications, they differ primarily in the degree to which computation remains resident on the QPU. Optimization-driven workflows are typically fragmented by frequent classical feedback, while time-evolution and linear-algebra workflows naturally support longer contiguous quantum compute regions. This observation motivates the locality-based comparison presented in the following section. A continually maintained application matrix and detailed pattern assignments are available online for reference and will evolve as additional hybrid quantum-classical workflows are analyzed.

\begin{table}[t]
\caption{Primary Hybrid Quantum--Classical Computational Patterns.}
\label{tab:patterns}
\centering
\renewcommand{\arraystretch}{1.15}
\begin{tabularx}{\columnwidth}{l X X c}
\hline
\textbf{Pattern} &
\textbf{QPU Function} &
\textbf{Representative Applications} &
\textbf{Locality} \\
\hline

QOOL & Objective-function evaluator & VQE, QAOA & Low \\
QASP & Subspace constructor & SKQD & Medium \\
QSMC & Sampling oracle & Quantum Monte Carlo & Medium \\
QTES & Time-evolution engine & Hamiltonian simulation & High \\
QLAO & Linear algebra solver & HHL, VQLS, QPE & High \\

\hline
\end{tabularx}
\end{table}

\section{Implications for Quantum Acceleration}

The proposed taxonomy reveals a consistent relationship between computational pattern and quantum execution locality. Optimization-driven workflows (QOOL), exemplified by widely studied variational algorithms such as VQE and QAOA \cite{cerezo2021variational}, constitute a major class of current NISQ applications which rely on frequent classical feedback and exhibit fragmented quantum execution. In contrast, subspace and sampling-based workflows (QASP and QSMC) naturally aggregate quantum work into batches, reducing synchronization overhead while retaining classical post-processing. Time-evolution and linear-algebra patterns (QTES and QLAO) support the highest execution locality by allowing computation to remain resident on the QPU across extended sequences of operations.

Higher quantum execution locality should not be interpreted
as uniformly beneficial in all regimes. In the NISQ era,
increasing locality by extending quantum-resident compute
regions also increases exposure to decoherence, accumulated
gate error, crosstalk, and readout noise. Thus, locality intro-
duces a tradeoff: longer contiguous quantum regions reduce
classical–quantum synchronization and orchestration over-
head, but may exceed the useful coherent execution window of
the device. The optimal dataflow is therefore not necessarily
the one with maximal locality, but the one that balances
quantum residency against hardware error rates, coherence
time, circuit depth, and error-mitigation or error-correction
overhead.

This relationship suggests that the effectiveness of quantum acceleration depends not only on algorithmic complexity, but also on workflow structure. High-locality patterns amortize classical-quantum synchronization over longer periods of quantum-resident computation, reducing orchestration overhead and making more effective use of quantum hardware. Conversely, fragmented workflows may remain limited by repeated interaction with classical systems, even when individual quantum kernels provide computational advantages.

This perspective encourages the reformulation of scientific workflows to increase quantum execution locality. Similar to the evolution of GPU computing \cite{nickolls2008cuda}, where performance improvements were achieved by restructuring applications around contiguous accelerator computation, future advances in hybrid quantum-classical computing may depend on designing workflows that maximize sustained quantum-resident execution.

\section{Hardware Implications: Pattern--Hardware Affinity}

An important implication of the proposed framework is that computational patterns are not hardware-neutral. Different hybrid dataflows place distinct demands on quantum processors, classical infrastructure, and system integration. Low-locality patterns such as QOOL emphasize rapid iteration and low-latency classical-quantum interaction, whereas high-locality patterns such as QTES and QLAO benefit from sustained quantum-resident execution, placing greater emphasis on coherence, gate fidelity, and extended circuit execution. Intermediate patterns, including QASP and QSMC, occupy a middle ground, favoring efficient batching and high-throughput sampling while relying on classical HPC resources for aggregation and reduced solves. These observations suggest that future hybrid systems may optimize for different classes of computational patterns rather than treating quantum processors as interchangeable accelerators. Evaluating these pattern-hardware affinities across diverse quantum computing platforms represents a natural next step and forms the basis of the empirical validation strategy outlined in the following section.

\section{Structural Characterization of Representative Hybrid Dataflows}
To provide preliminary quantitative grounding for the proposed framework, Table~\ref{tab:workflow_locality_characterization} summarizes representative locality characteristics across several common hybrid quantum-classical dataflows. Rather than focusing on hardware-specific performance metrics, this characterization emphasizes structural properties of the dataflow itself, including synchronization frequency, quantum residency, the nature of the quantum compute region, and locality. These quantities provide a practical means of comparing dataflows across application domains while remaining largely independent of a particular hardware platform or implementation. Consistent with the preceding analysis, optimization-driven workflows occupy the low-locality regime, while time-evolution and linear-algebra-based approaches exhibit substantially higher quantum residency. Although qualitative in nature, this characterization demonstrates that the proposed computational patterns correspond to observable workflow behaviors and provides a foundation for future empirical studies of hybrid quantum execution.

The framework proposed in this work is intended to be evaluated through a two-phase validation strategy. The first phase focuses on structural characterization, using representative hybrid workflows to estimate synchronization frequency, relative quantum residency, and execution locality from existing algorithmic descriptions and published performance data. This provides an initial qualitative and order-of-magnitude assessment of how different computational patterns map to the proposed locality framework while remaining largely independent of a particular hardware platform.

The second phase will provide empirical validation by implementing representative kernels for the four selected benchmark workflows (QOOL, QASP, QSMC, and QTES) within a common hybrid benchmarking framework such as the Quantum Metric Assessment Platform (QMAP) \cite{QMAPRepo}. These minimal instantiations will enable detailed diagnostic profiling of execution behavior, including synchronization frequency, communication overhead, quantum residency, and overall workflow performance across heterogeneous quantum-classical platforms. Together, these two phases establish both a conceptual methodology for characterizing hybrid computational patterns and a practical path toward quantitative evaluation and benchmarking. QLAO is omitted from the initial implementation because representative linear-algebra benchmarks and implementations are less mature within the current framework, but remain a natural extension of the validation methodology.

\begin{table*}[t]
\centering
\caption{Illustrative locality characteristics of representative hybrid quantum--classical workflows. Values are qualitative and intended to capture relative execution behavior rather than empirical measurements. Relative quantum residency refers to the duration over which computation remains resident on the QPU without requiring host intervention or classical synchronization.}
\label{tab:workflow_locality_characterization}

\begin{tabular}{l l l l l l}
\toprule
\textbf{Workflow} &
\textbf{Pattern} &
\textbf{Host--QPU Crossings} &
\textbf{Relative Quantum Residency} &
\textbf{Dominant Dataflow} &
\textbf{Locality Class} \\
\midrule

VQE &
QOOL &
Very High $\mathcal{O}(10000)$ &
Low &
Expectation evaluation &
Fragmented \\

QAOA &
QOOL &
High $\mathcal{O}(1000)$ &
Low &
Objective evaluation &
Fragmented \\

SKQD &
QASP &
Medium $\mathcal{O}(100)$ &
Medium &
Subspace construction batch &
Batched \\

Quantum Sampling &
QSMC &
Medium $\mathcal{O}(100)$ &
Medium &
Sampling batch &
Batched \\

Hamiltonian Simulation &
QTES &
Low $\mathcal{O}(1)$ &
High &
Time evolution trajectory &
Contiguous \\

Quantum Linear Solver &
QLAO &
Low $\mathcal{O}(1)$ &
High &
Linear algebra solve &
Contiguous \\

\bottomrule
\end{tabular}

\end{table*}

\section{Reformulating Scientific Workflows}

The preceding analysis suggests that the effectiveness of quantum acceleration depends as much on dataflow structure as on algorithmic capability. We examine how existing scientific workflows can be reformulated to increase quantum execution locality and reduce classical-quantum synchronization.

Increasing locality requires identifying portions of a workflow that can remain resident on the QPU across multiple computational stages before measurement. The value of intermediate quantum computation lies not in recovering classical values, but in preserving quantum state across successive coherent transformations prior to measurement. Achieving this may involve batching quantum work to reduce synchronization overhead, restructuring workflows around subspace or sampling representations, and composing larger quantum-resident regions using linear algebra, time evolution, and other quantum primitives. Such reformulations seek to maximize contiguous quantum execution while minimizing repeated interaction with classical resources.

Taken together, these observations suggest a shift from viewing quantum processors as peripheral accelerators toward treating them as computational domains in which suitably structured portions of a workflow can reside. The challenge is therefore not only identifying quantum-suitable problems, but identifying which workflows can be transformed to better align with the structural properties of quantum hardware.

\section{Future Directions}

The framework presented in this work suggests several directions for future research. A central challenge is increasing quantum execution locality by transforming fragmented hybrid workflows into dataflows that support more sustained quantum-resident computation through batching, deferred measurement, and workflow reformulation. More broadly, this raises the question of \textit{workflow transformability}: which scientific applications can be systematically reorganized to expose larger contiguous quantum compute regions while preserving computational correctness and scientific utility?

The structural characterization and two-phase validation methodology outlined in this work provide a foundation for evaluating both the computational patterns across heterogeneous quantum-classical platforms and the relationships between workflow structure, hardware capabilities, and performance. The first phase characterizes representative workflows using structural metrics and order-of-magnitude timing estimates, while the second implements representative benchmark kernels for diagnostic profiling and performance evaluation. Together, these phases establish a practical methodology for testing the proposed taxonomy and investigating pattern-hardware affinity across hybrid quantum-classical systems.

As quantum hardware continues to mature, this locality-oriented perspective may help guide the co-design of algorithms, runtime systems, and hybrid computing architectures, providing a systematic framework for understanding how workflow structure influences the realization of quantum acceleration.

\section{Conclusion}

This work introduces the \emph{Quantum Execution Locality Framework (QELF)}, a dataflow-oriented framework for understanding hybrid quantum-classical scientific computing through recurring computational patterns and quantum execution locality. By classifying representative hybrid workflows according to the computational role of the QPU, QELF provides a common structural abstraction for analyzing execution behavior across diverse applications.

The proposed framework suggests that the effectiveness of quantum acceleration depends not only on algorithmic capability, but also on the extent to which computation remains resident on the QPU. By combining computational patterns, execution locality, and a two-phase validation strategy, QELF establishes a practical foundation for future benchmarking, hardware-software co-design, and the continued evolution of hybrid quantum-classical systems.

Ultimately, the question is not only which problems quantum computers can solve, but how those problems are expressed.

\bibliographystyle{unsrt}
\bibliography{bibliography}

@article{peruzzo2014variational,
  title   = {A variational eigenvalue solver on a quantum processor},
  author  = {Peruzzo, Alberto and McClean, Jarrod and Shadbolt, Peter and Yung, Man-Hong and Zhou, Xiao-Qi and Love, Peter J. and Aspuru-Guzik, Al{\'a}n and O'Brien, Jeremy L.},
  journal = {Nature Communications},
  volume  = {5},
  pages   = {4213},
  year    = {2014},
  doi     = {10.1038/ncomms5213},
  url     = {https://doi.org/10.1038/ncomms5213},
  archivePrefix = {arXiv},
  eprint  = {1304.3061},
  primaryClass = {quant-ph}
}

@article{yu2025skqd,
  title         = {Quantum-Centric Algorithm for Sample-Based Krylov Diagonalization},
  author        = {Yu, Jeffery and Robledo-Moreno, Javier and Iosue, Joseph T. and Bertels, Luke and Claudino, Daniel and Fuller, Bryce and Groszkowski, Peter and Humble, Travis S. and Jurcevic, Petar and Kirby, William and others},
  journal       = {arXiv preprint arXiv:2501.09702},
  year          = {2025},
  archivePrefix = {arXiv},
  eprint        = {2501.09702},
  primaryClass  = {quant-ph},
  url           = {https://arxiv.org/abs/2501.09702},
  doi           = {10.48550/arXiv.2501.09702}
}

@article{cerezo2021variational,
  author  = {M. Cerezo and others},
  title   = {Variational Quantum Algorithms},
  journal = {Nature Reviews Physics},
  volume  = {3},
  pages   = {625--644},
  year    = {2021},
  doi     = {10.1038/s42254-021-00348-9}
}

@misc{QMAPRepo,
  author = {Landfield, Ryan and Dayal, Arnav and Patel, Ria},
  title = {{QMAP: Quantum Metric Assessment Platform}},
  year = {2026},
  howpublished = {\url{https://github.com/AstroLando/QMAP}},
  note = {GitHub repository}
}

@article{harrow2009hhl,
  author    = {Aram W. Harrow and Avinatan Hassidim and Seth Lloyd},
  title     = {Quantum Algorithm for Linear Systems of Equations},
  journal   = {Physical Review Letters},
  volume    = {103},
  number    = {15},
  pages     = {150502},
  year      = {2009},
  doi       = {10.1103/PhysRevLett.103.150502},
  url       = {https://doi.org/10.1103/PhysRevLett.103.150502}
}

@article{lloyd1996universal,
  author  = {Seth Lloyd},
  title   = {Universal Quantum Simulators},
  journal = {Science},
  volume  = {273},
  number  = {5278},
  pages   = {1073--1078},
  year    = {1996},
  doi     = {10.1126/science.273.5278.1073},
  url     = {https://doi.org/10.1126/science.273.5278.1073}
}

@article{montanaro2015montecarlo,
  author  = {Ashley Montanaro},
  title   = {Quantum Speedup of Monte Carlo Methods},
  journal = {Proceedings of the Royal Society A},
  volume  = {471},
  number  = {2181},
  pages   = {20150301},
  year    = {2015},
  doi     = {10.1098/rspa.2015.0301},
  url     = {https://doi.org/10.1098/rspa.2015.0301}
}

@article{brassard2002amplitude,
  author  = {Gilles Brassard and Peter H{\o}yer and Michele Mosca and Alain Tapp},
  title   = {Quantum Amplitude Amplification and Estimation},
  journal = {Contemporary Mathematics},
  volume  = {305},
  pages   = {53--74},
  year    = {2002},
  doi     = {10.1090/conm/305/05215},
  url     = {https://doi.org/10.1090/conm/305/05215}
}

@article{nickolls2008cuda,
  author  = {John Nickolls and Ian Buck and Michael Garland and Kevin Skadron},
  title   = {Scalable Parallel Programming with CUDA},
  journal = {Queue},
  volume  = {6},
  number  = {2},
  pages   = {40--53},
  year    = {2008},
  doi     = {10.1145/1365490.1365500}
}

@article{Beck2024IntegratingQuantumHPC,
  author       = {Beck, Thomas and Baroni, Alessandro and Bennink, Ryan and Buchs, Gilles and Coello Pérez, Eduardo Antonio and Eisenbach, Markus and Ferreira da Silva, Rafael and Gopalakrishnan Meena, Muralikrishnan and Gottiparthi, Kalyan and Groszkowski, Peter and Humble, Travis S. and Landfield, Ryan and Maheshwari, Ketan and Oral, Sarp and Sandoval, Michael A. and Shehata, Amir and Suh, In-Saeng and Zimmer, Christopher},
  title        = {Integrating Quantum Computing Resources into Scientific HPC Ecosystems},
  journal      = {arXiv preprint arXiv:2408.16159},
  year         = {2024},
  note         = {Preprint; submitted 28 Aug 2024, \url{https://doi.org/10.48550/arXiv.2408.16159}},
}
\end{document}